\documentclass[a4paper,11pt]{article}
\usepackage{jinstpub} % Standard JINST class for formatting
\usepackage{lineno}
\usepackage{graphicx}
\usepackage{float} % Used for the [H] placement option
\usepackage{setspace}
\usepackage{tabularx} % For auto-adjusting column widths
\usepackage{booktabs} % Essential for professional tables
\usepackage{url}
\usepackage[colorlinks=true, urlcolor=blue, linkcolor=blue, citecolor=blue]{hyperref}
\usepackage{doi}

\title{Stacked geometry concept for CosmicWatch based muon detectors in coincidence mode}
\author[a]{Ishannita Banerjee,}
\author[b]{Swagato Banerjee$,^{1,}$\note{Corresponding author.}}
\author[a]{Douglas Jackson,}
\author[a]{John Naber}

\affiliation[a]{Department of Electrical Engineering, University of Louisville, Louisville, KY 40292, USA.}
\affiliation[b]{Department of Physics and Astronomy, University of Louisville, Louisville, KY 40292, USA.}

\emailAdd{swagato.banerjee@louisville.edu}

\abstract{
CosmicWatch-based muon detectors are inexpensive, handheld, battery-powered, portable instruments designed to measure the flux of secondary muons produced when primary cosmic rays continuously strike the Earth's atmosphere. 
As these muons pass through the detector, plastic scintillators generate light, which is collected by silicon photomultiplier (SiPM) sensors and converted to an electric signal. 
The signal is processed by an Arduino-based acquisition system and recorded as count rates corresponding to the cosmic muon flux. 
A stacked configuration consisting of two scintillator modules wrapped together and read by two SiPM sensors, which are operated in coincidence mode via linked Arduino microcontrollers, is presented.
The novelty of this work is in demonstrating that detector geometry improves signal purity by defining a constrained solid angle and enabling coincidence filtering, thereby reducing uncorrelated background and local electronic noise. 
The system was deployed during the 2024 total solar eclipse, demonstrating stable field operation and reliable flux monitoring. 
The mechanical design, coincidence logic implementation, and performance characterization of the instrument are described. 
Overall, this work demonstrates a scalable, robust, and inexpensive architecture suitable for geographically distributed muon flux monitoring arrays.
}


\begin{document}
\maketitle
\flushbottom

\section{Introduction}
\label{sec:intro}

Cosmic rays are a stream of fast-moving particles that continuously strike the Earth's atmosphere from the interstellar medium.  
Galactic sources, such as supernova remnants, generate a relatively steady stream of high-energy primary cosmic rays with energies spanning many orders of magnitude.  
Turbulence in the Sun's chromosphere generates a flux of lower-energy charged particles, known as the solar wind, whose flux is significantly modulated by solar activity~\cite{PDG2024}.

While the composition of these primary cosmic rays is predominantly protons and some alpha particles, these hadrons rarely reach the Earth's surface intact.
Instead, they interact with the atomic nuclei of oxygen and nitrogen in the upper atmosphere to initiate extensive air showers.
These collisions generate cascades of secondary particles, primarily pions and kaons, which are unstable and decay rapidly.
It is the decay of these pions and kaons that produces atmospheric muons, the principal constituent of secondary cosmic radiation observed at sea level.
Due to their high penetrating power and relativistic lifetimes, these cosmic muons provide a robust, high-rate observable flux for investigating astrophysical phenomena, space weather, and detector instrumentation physics for diverse applications such as muon tomography~\cite{Grieder2001,Yang2019,Lincoln:2023wkb}.

The landscape of cosmic muon flux detection has significantly evolved since the advent of the CosmicWatch v1 desktop muon detectors, which demonstrated that low-cost silicon photomultipliers (SiPMs) offer a viable alternative to traditional bulky photomultiplier tubes (PMTs) for portable applications~\cite{Axani2017}. 
This was followed by the CosmicWatch v2 detector, which consolidated open-source hardware and software into a standardized platform that became widely adopted for outreach activities~\cite{Axani2018}. 
Most recently, the release of CosmicWatch v3 detectors introduced significant hardware and firmware enhancements, improving sensitivity and data acquisition capabilities~\cite{Axani2025}. 

The stacked geometry concept described here is based on the established CosmicWatch v2 design. Our aim is not to revise the overall platform, but to present a specialized geometry configuration. By arranging plastic scintillators in a compact, stacked configuration, this geometry defines a precise solid angle for incident particles, improving cosmic-ray flux measurements compared to single-layer, open-acceptance designs.

The focus of this work is the optimization of coincidence logic and data acquisition parameters to maximize signal purity while preserving the portability of the CosmicWatch detector. We characterize key instrumental factors, including effective detection area, silicon photomultiplier dark-noise suppression, and electronic dead time. This approach reduces systematic uncertainties and provides the stability required for precision monitoring.

To validate this configuration, we deployed the stacked geometry configuration of a self-assembled CosmicWatch-based muon detector during the 2024 total solar eclipse, which served as a rigorous environmental stress test. The resulting data were used to evaluate timing precision and thermal stability, demonstrating the detector's robustness for small-scale astrophysical research.

\section{Detector Design and Coincidence Principle}
\label{sec:design}

\subsection{Standard Components and Coincidence Operation}

\begin{figure}[!hbtp]
    \centering
    \fbox{\includegraphics[width=0.95\linewidth]{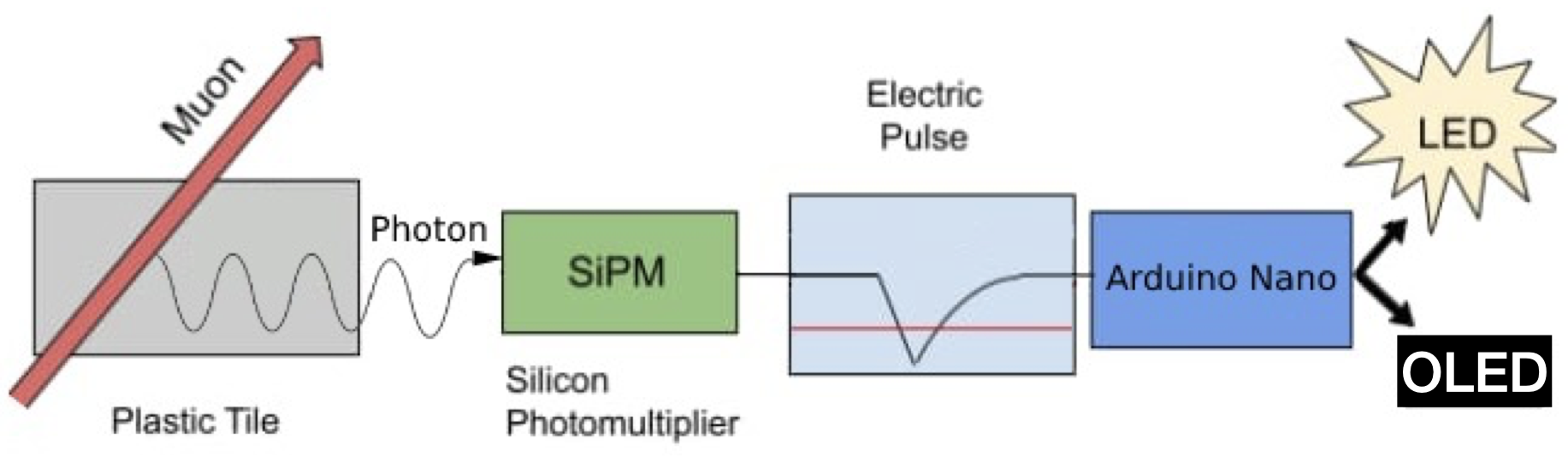}}
    \caption{General working principle of a CosmicWatch-based cosmic muon detector.}
    \label{fig:muon_detection_flow}
\end{figure}

Each CosmicWatch-based cosmic muon detector module~\cite{Axani2018} is built locally from commercially available open-source components for cost-effectiveness and reproducibility. It consists of a $5 \times 5 \times 1 \, \text{cm}^3$ BC408 plastic scintillator tile produced by Bicron, which emits light when a muon passes through. This light is detected by a $6 \, \text{mm} \times 6 \, \text{mm}$ photodiode (MICROFC-60035-SMT-TR1 by ONSEMI) used as a silicon photomultiplier (SiPM). The scintillator is wrapped in aluminium foil and coupled to the SiPM with optical gel. The SiPM converts the light into an electric pulse, which is sent to an Arduino Nano board for amplification. Signals above threshold flash an LED, and the count rate is shown on the OLED panel. A schematic of this functionality is shown in Figure~\ref{fig:muon_detection_flow}.

To suppress electronic noise and false counts, two such modules can be operated in a coincidence mode. This mode requires a pulse to be registered simultaneously (within a 0.1-second window) by two separate detectors, which are interconnected via a short coincidence cable. This ensures that penetrating secondary muons, and not local electronic noise, are registered. 

\subsection{Stacked Geometry and Performance Optimization}
\label{sec:sandwich}

\begin{figure}[!ht]
\centering
\fbox{\includegraphics[height=0.25\textheight,width=0.45\textwidth]{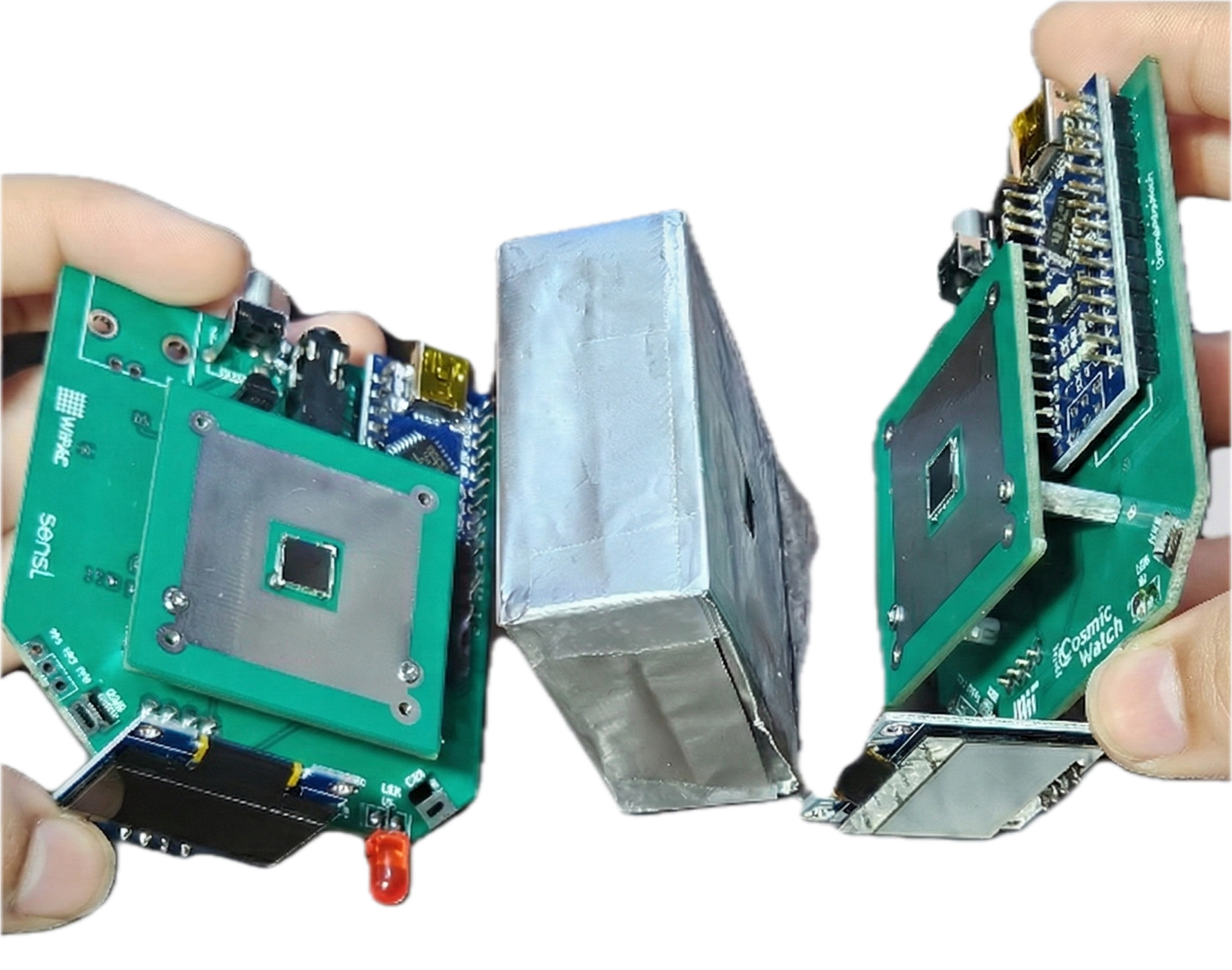}}
\fbox{\includegraphics[height=.25\textheight,width=0.45\textwidth]{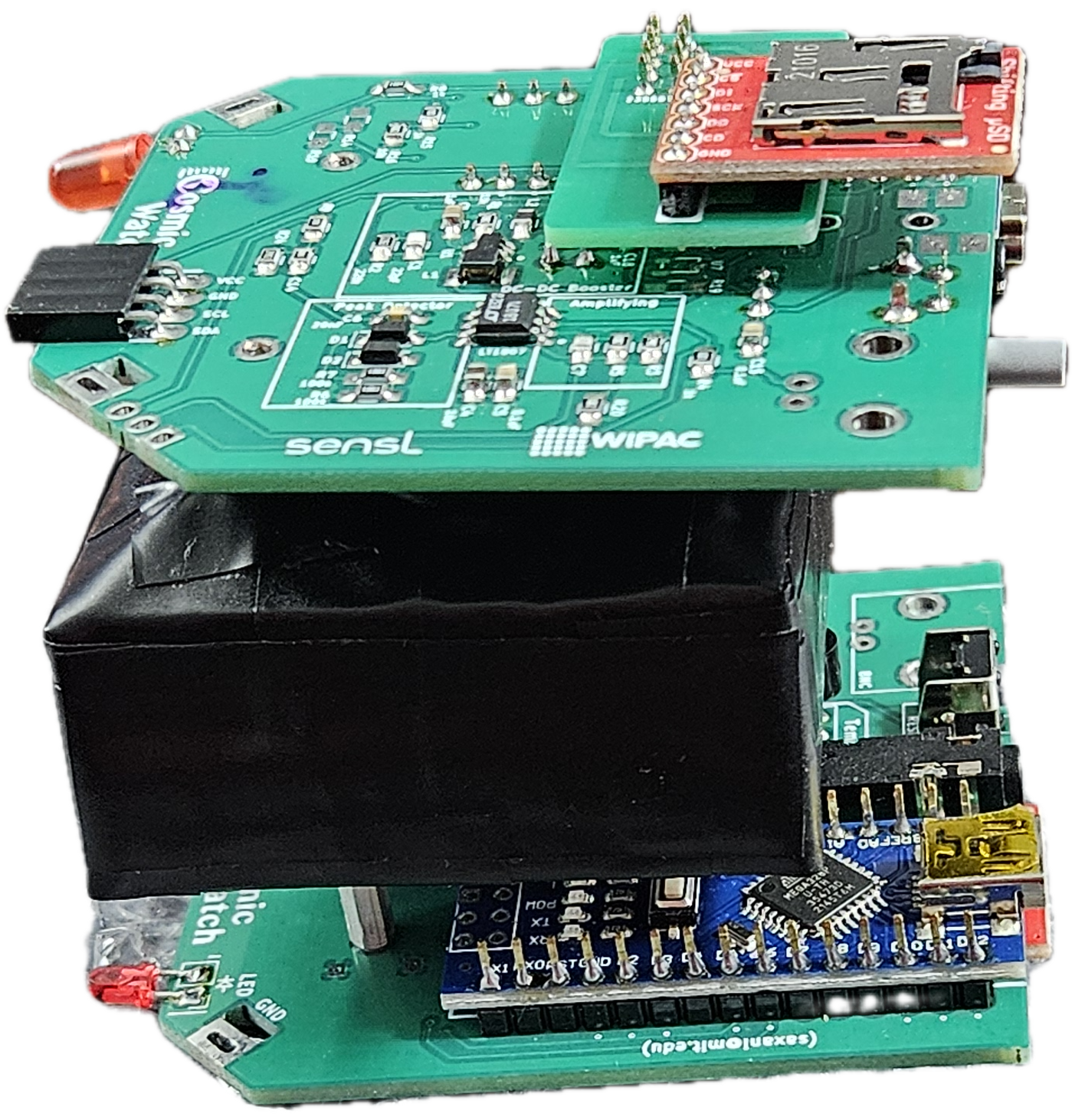}}
\caption{Left: Two plastic scintillators are stacked and wrapped in a single aluminum foil, with holes cut for the two SiPMs. Right: The final configuration where two SiPM boards are aligned on either side of the stacked scintillators, forming a compact, doubled-active-volume unit.}
\label{fig:sandwich_assembly}
\end{figure}

The construction involves stacking two plastic scintillator tiles on top of each other, wrapping the stack in a single aluminum foil layer with holes cut on either side to align with the SiPM chips, aligning two separate SiPM chips on opposite sides of the stack so that the scintillators are sandwiched between them, and interconnecting the two Arduino Nanos via the coincidence cable. 
This arrangement ensures that a single muon passing through the entire assembly is likely to register a signal in both scintillators, and thus both SiPMs, creating a high-fidelity coincidence count.
This stacked geometry concept, aimed at increasing detection efficiency and reducing accidentals in the statistical count rate in a coincidence-mode detection, is presented in Figure~\ref{fig:sandwich_assembly}.

\section{Performance Characterization}
\label{sec:performance}

\begin{figure}[!htbp]
\centering
\fbox{\includegraphics[width=0.85\textwidth]{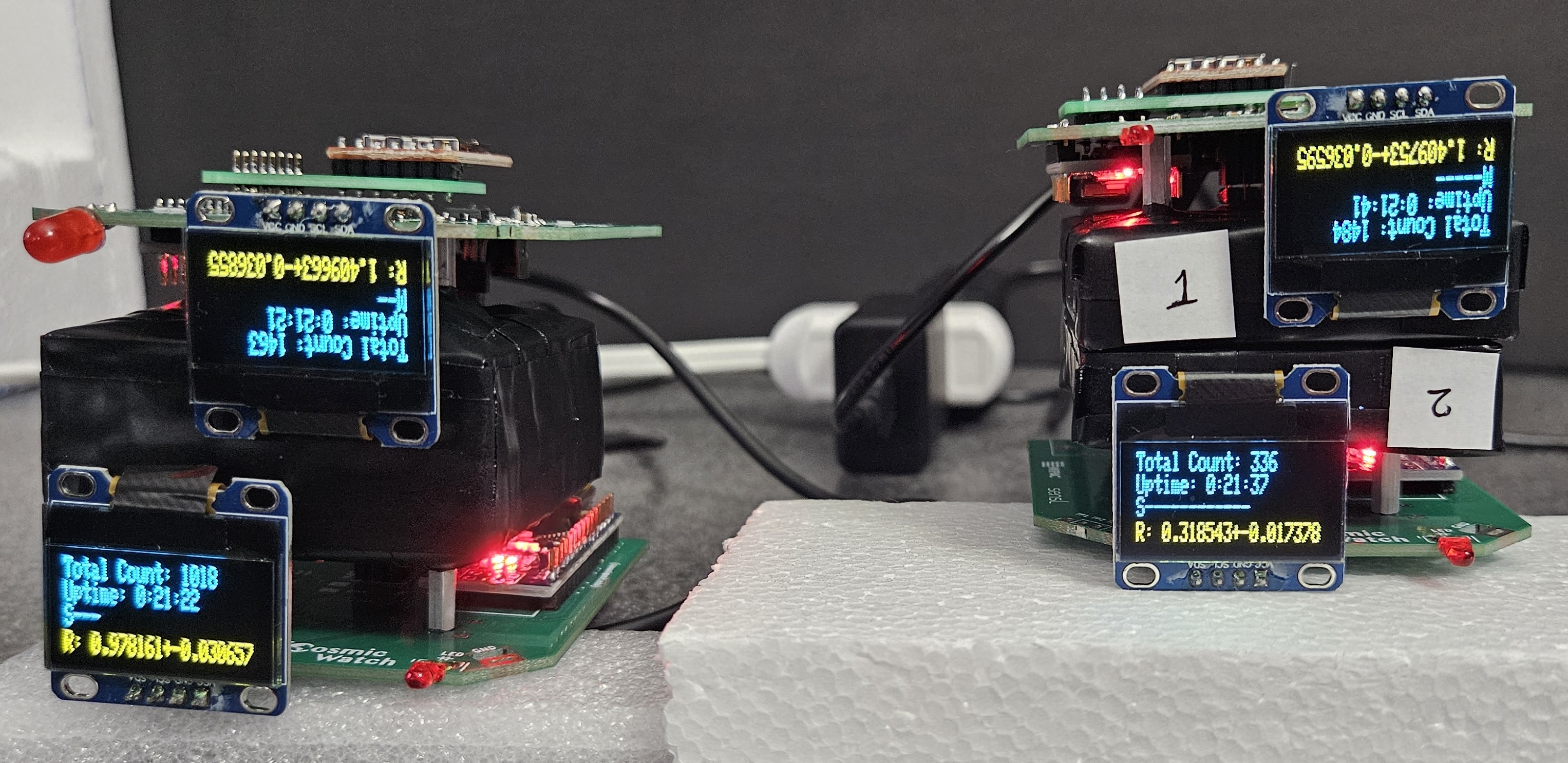}}
\caption{Comparison of the original, separated coincidence detectors (right) and the stacked detector (left).}
\label{fig:count_rate_comparison}
\end{figure}

To validate the improved statistical performance of the stacked detector, its count rate  was directly compared against the original, spatially separated coincidence detector. 
As shown in Figure~\ref{fig:count_rate_comparison} for the separated and optimized detector geometry, the measured count rates ($R$) were:
\begin{itemize}
    \item Original Coincidence Detector (Separated): $R_{\text{orig}} = (0.32 \pm 0.02)$ counts per second.
    \item Stacked Detector (Optimized): $R_{\text{sand}} = (0.98 \pm 0.03)$ counts per second.
\end{itemize}

The stacked detector achieved a count rate increase of:
$$
\frac{R_{\text{sand}}}{R_{\text{orig}}} = \frac{0.98}{0.32} \approx 3.06
$$
The new configuration increased the statistical count rate by about a factor of three, achieving the goal of high-rate detection critical for short-duration studies. 
According to the manual~\cite{Onsemi_CSeries}, the
MICROFC-60035-SMT-TR1 chip exhibits stable performance over the range $-40^\circ\mathrm{F}$ to $185^\circ\mathrm{F}$.
We verified that our measurements remain stable at the percent level by repeating the experiments at different times of day, where the  ambient temperature varied over a range of $60^\circ\mathrm{F}$ to $75^\circ\mathrm{F}$.

While stacked geometry had also been studied in Ref.~\cite{Nakamori:2023ebe}, we present the first direct comparison of the performance characterization based on count rate in the same ambient environment and with both configurations operating in tandem.

\section{Performance Testing during 2024 Total Solar Eclipse}
\label{sec:validation}

\subsection{Experimental results} 

\begin{figure}[!htbp]
\centering
\includegraphics[width=0.8\textwidth]{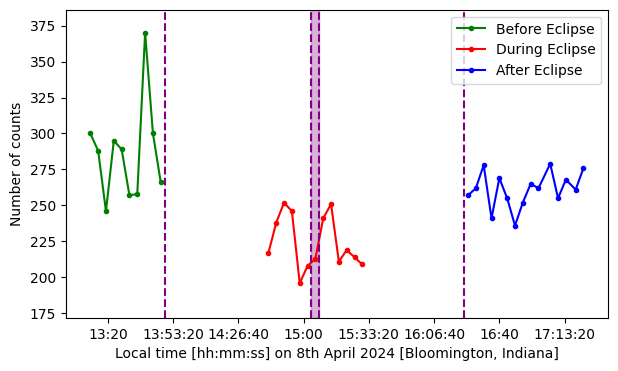}
\caption{Poisson fit to the baseline counts at Bloomington, IN. The baseline mean is $\lambda=535.8$. The measured counts during the eclipse ($x=414$) are highly statistically significant, highlighting the high performance of the detector.}
\label{fig:timeseries_bloomington}
\end{figure}

The stacked detector was used to measure the time-dependent flux of secondary cosmic rays during the total solar eclipse event on April 8, 2024, at Bloomington, Indiana. 
Measurements were taken in fixed 4-minute intervals spanning the period before, during, and after maximal coverage, as shown in Figure~\ref{fig:timeseries_bloomington}. 
Baseline counts (before and after the eclipse) were fitted with a Poisson distribution function, establishing detector stability and yielding mean expected count rates of $\lambda = 271.4$ and $267.3$ with and without an outlier count of 370, respectively. A similar fit to the eclipse counts yields a mean rate of $x = 224.3$, corresponding to reductions of 17.4\% and 16.1\% from the baseline, with and without the outlier, respectively.

\begin{table}[ht]
\centering

\begin{tabular}{llcl}
\toprule
\textbf{Time} & \textbf{Phase} & \textbf{Temp} & \textbf{Observational notes} \\
\textbf{(EDT)} & & \textbf{(°F)} & \\
\midrule
1:30 PM & pre-eclipse baseline & 73 & baseline atmospheric conditions \\
1:49 PM & partial eclipse onset & 72 & initiation of solar occultation \\
2:15 PM & intermediate partial phase & 70 & reduction in insolation; shadow sharpening \\
2:45 PM & deep partial phase & 66 & ambient cooling observed \\
3:00 PM & pre-totality transition & 64 & onset of nocturnal biological behaviors \\
3:04 PM & totality onset & 62 & minimum temperature; solar corona visible \\
3:06 PM & maximum eclipse & 62 & peak obscuration achieved \\
3:08 PM & totality conclusion & 63 & photospheric re-emergence; diamond ring effect \\
3:30 PM & post-totality recovery & 67 & rapid thermal rebound detected \\
4:00 PM & late partial phase & 70 & return to nominal daylight luminosity \\
4:22 PM & partial eclipse conclusion & 71 & termination of eclipse event \\
\bottomrule
\end{tabular}
\caption{Chronological temperature progression and observational notes during the 2024 total solar eclipse at Bloomington, IN~\cite{IGWS_Eclipse2024}.}
\label{tab:eclipse_observations}
\end{table}

\subsection{Interpretation of the data}
The Sun continuously releases a “solar wind” composed of protons, but these particles generally do not carry enough energy to initiate reactions that generate enough muons to reach sea level and produce a continuous flux of cosmic muons.
For a muon to be created high in the atmosphere and still survive the roughly 15 km journey to the ground, the incoming primary particle must have an energy well above a few GeV.
Consequently, most muons measured at sea level are produced by galactic cosmic rays arriving nearly uniformly from all directions. The Moon subtends only about 0.2 square degrees, corresponding to roughly 0.001\% of the detector’s visible sky, even if it blocked 100\% of the cosmic rays coming from behind it.

The most likely interpretation of the observed “dip” is the temperature change reported in Table~\ref{tab:eclipse_observations}~\cite{IGWS_Eclipse2024}, which links these reductions not to the Moon blocking the particles, but to atmospheric cooling. During an eclipse, the air temperature decreases, causing the atmosphere to contract and become more dense. This shifts the altitude at which muons are generated and influences their chance of reaching the ground, leading to a “production-layer lowering” scenario in which a cooler, compressed atmosphere reduces the muon path length and alters the measured flux. Moreover, the detector also records some dark or electronic noise in addition to real muon events. As the air cools during the eclipse, this noise component declines, further lowering the total count rate.

These findings confirm that the optimized, stacked detector offers sufficient statistical precision and sensitivity to clearly detect subtle, short-lived variations in the cosmic-ray flux.

\section{Conclusion and Outlook}
\label{sec:conclusion}

This report presents a significant technical improvement to the widely adopted CosmicWatch-based muon detection system. The stacked geometry successfully achieved high-rate coincidence detection, increasing the count rate by over a factor of three. 
This technical achievement directly addresses the statistical limitations of previous similar experiments, making it suitable for high-precision, time-dependent studies.

The deployment of the optimized detector during the 2024 solar eclipses served as a rigorous performance test. The detector demonstrated excellent stability and provided statistically robust measurements of cosmic ray flux attenuation, which can be utilized during future total solar eclipses.

This work presents a proven, low-cost, and highly reproducible detector design for the wider scientific community that harnesses the natural flux of cosmic muons to reveal the internal structure of large-scale objects. This capability enables diverse applications for nuclear security, civil engineering, geophysics, archaeology, atmospheric science, and industrial inspection.

\appendix
\section{Acknowledgments}
The researchers gratefully acknowledge Dr. Spencer Axani for discussions about the CosmicWatch detector~\cite{Axani2017,Axani2018,Axani2025}, and the NASA Techrise Team for the donation of the plastic scintillator tiles~\cite{NASATechRise_Muons}.

\bibliographystyle{JHEP}
\bibliography{references}

\end{document}